\documentclass[showpacs,twocolumn,superscriptaddress,aps]{revtex4}%
\usepackage{amsmath}
\usepackage{graphicx}
\usepackage{amsfonts}
\usepackage{amssymb}%
\begin{document}
\title{Processing multi-photon state through operation on single photon: methods and applications}
\author{Qing Lin}
\email{qlin@mail.ustc.edu.cn}
\affiliation{College of Information Science and Engineering, Huaqiao University (Xiamen),
Xiamen 361021, China}
\author{Bing He}
\email{bhe98@earthlink.net }
\affiliation{Institute for Quantum Information Science, University of Calgary, Alberta T2N
1N4, Canada}
\author{J\'{a}nos A. Bergou}
\affiliation{Department of Physics and Astronomy, Hunter College of
the City University of New York, 695 Park Avenue, New York, New York
10065, USA}
\author{Yuhang Ren}
\affiliation{Department of Physics and Astronomy, Hunter College of
the City University of New York, 695 Park Avenue, New York, New York
10065, USA}


\begin{abstract}
Multi-photon states are widely applied in quantum information technology. By the methods presented in this paper, the structure of a multi-photon state in the form of multiple single photon qubit product can be mapped to a single photon qudit, which could also be in separable product with
other photons. This makes the possible manipulation of such multi-photon states in the way of processing single photon states. The optical realization of unknown qubit discrimination [B. He, J. A. Bergou, and Y.-H. Ren, Phys. Rev. A 76, 032301 (2007)] is simplified with the transformation methods. Another application is the construction of quantum logic gates, where the inverse transformations back to the input state spaces are also necessary. We especially show that the modified setups to implement the transformations can realize the deterministic multi-control gates (including Toffoli gate) operating directly on the products of single photon qubits.
\end{abstract}
\maketitle

\section{\bigskip Introduction}

Photonic states are suitable to carrying quantum information for their flexibility and robustness against decoherence
effects. The flying qubits in various quantum computing schemes are encoded in photonic states \cite{P}, and many tasks in quantum cryptography
and quantum communications are performed with the aid of photonic states as well \cite{cryp, repeater}. In most of protocols, an optical system
should process the input of multi-photon state, which is often a tensor product of the signal plus the ancilla. In the scheme of unknown quantum state discrimination \cite{Bergou}, for example, the inputs should be prepared as the tensor product of the program and
the data state, $|\psi_1\rangle|\psi_2\rangle|\psi_3\rangle$, where the qubit $|\psi_i\rangle$ ($i=1,2$ for the program and $i=3$ for the data) 
could be an unknown linear combination of the polarization modes $|H\rangle$ and $|V\rangle$ ($H$ represents
horizontal and $V$ vertical polarization) of a single photon. When a multi-photon input state is processed by a linear optical device together with the post-selection through measurement on part of the output, there will be an upperbound for the success probability to obtain the desired target state \cite{efficiency}. Non-deterministic operations are inevitable in such systems, so it demands more resources to realize a processing task definitely.

The situation will be different when we deal with single photon states. Quantum information can be encoded in spatial degree of freedom 
(different spatial paths) and internal degree of freedom (polarization, etc.) of a single photon.
A significant advantage in using single photon states of multiple spatial modes is that arbitrary unitary operation on such states can be deterministically realized with a linear optical device \cite{Reck}. This result can be generalized to realize arbitrary positive-operator-valued-measurement (POVM) on multiple path-mode single photon states \cite{h-b-w}. For instance, it will be possible to
perform the above-mentioned unknown qubits discrimination, if one could realize a map
\begin{align}
|\psi_1\rangle|\psi_2\rangle|\psi_3\rangle&=c_1|HHH\rangle+c_2|HHV\rangle+\cdots+c_8|VVV\rangle\nonumber\\
&\rightarrow c_1|1\rangle+c_2|2\rangle+\cdots+c_8|8\rangle
\label{1}
\end{align}
to a corresponding multi-mode single photon inheriting the coefficients $c_i$ of the input multi-photon state ($|1\rangle$, $\cdots$, $|8\rangle$ are the spatial modes here). In particular, the unambiguous discrimination of any pair of multi-mode single photon states and of even more such states can be simply realized by linear optics \cite{usd}.

In this paper we discuss the methods to realize such transformations and some of their useful applications. It has been proposed that the transformation in Eq. (\ref{1}) could be realized by teleportation \cite{parity}. However, the method can be simplified much further if we improve
on the necessary circuits. Here we also present a purely circuit-based approach without teleportation to realize the transformations in the form
\begin{align}
&~~~~c_1|HHH\rangle+c_2|HHV\rangle+\cdots+c_8|VVV\rangle\nonumber\\
&\rightarrow |+\rangle_1|+\rangle_2(c_1|1\rangle+c_2|2\rangle+\cdots+c_8|8\rangle)_3,
\label{20}
\end{align}
in which the third photon inheriting the coefficients $c_i$ of the input multi-photon state will be in tensor product with two other photons of the state $|+\rangle=\frac{1}{\sqrt{2}}(|H\rangle+|V\rangle)$. This kind of transformations can be realized without ancilla and with a quantity of circuit resources scaling linearly to the photon number in input states.

An important application of these transformations is the construction of quantum networks or quantum logic gates.
As there is no direct interaction between photonic states, the nonlinear media of strong intensity seem necessary to realize photonic logic
gates. However, the best cross-Kerr nonlinearity created by electromagnetically induced transparency (EIT) techniques \cite{EIT} thus far is still
in the range of weak nonlinearity. The development of photonic logic gates is therefore mainly along the line of the Knill-Laflmme-Milburn
(KLM) protocol \cite{KLM}, which applies linear optical circuit, entangled ancilla state and photon number resolving detection to realize a two-qibit gate probabilistically. The KLM protocol and its improvements within linear optics are all probabilistic, if the realistic resources are available \cite{P}. Since the input states for many key logic gates or quantum networks can be simply encoded as the initial states in Eq. (\ref{1}) and (\ref{20}), it is reasonable to first convert them to the corresponding single photon states and then use a linear optical circuit to deterministically implement the unitary operations for the gates. The remaining work will be converting the processed states in multiple path modes back to polarization space. The transformations to single photon states and their inverse processes are realizable using weak nonlinearity, linear optics, as well as photon number resolving detection which could be performed by various methods. This idea of realizing a quantum network is different from the conventional method of decomposing it into the product of two-qubit and one-qubit gates \cite{Nielsen}.

Feasible though the realization of all kinds of deterministic two-qubit, triple-qubit gates, etc, in the approach is, it is impractical to implement
a quantum network involving large number of input qubits this way, as the number of the spatial modes carried by the transformed single photon is exponentially large ($2^n$ for $n$ input qubits). For a special type of multi-qubit gates, however, the necessary resources could be much fewer than those required by the conventional decomposition method into two-qubit gates, if we modify an element gate used for the previously mentioned transformations to single photon states a bit. This type of multi-qubit gates include Toffoli gate \cite{Toffoli}, a gate key to many quantum information processing tasks. We will discuss the building of these gates in detail.

The rest of the paper is organized as follows. In Sec. \ref{sec2}, we discuss the designs of the gates which are related to the teleportation of multi-photon states to the corresponding single photon states. The improvement on the teleportation strategy in \cite{parity} is given in \ref{tele}.
The purely circuit-base approach to realizing the map in Eq. (\ref{20}) is presented in Sec. \ref{sec3}, where the steps to transform input two-photon and triple-photon states are respectively illustrated in details. The application of the transformation methods in constructing quantum networks is discussed in Sec. \ref{sec4}; in this section we discuss the realization of general two-qubit gate, general multi-qubit gate and especially multi-control gates including Toffoli gate, respectively. Finally a few remarks conclude the paper in the last section.

\section{Teleportation approach}
 \label{sec2}

Firstly, we discuss the approach to transform multi-photon states to the corresponding single photon states through teleportation.
The designs of the necessary gates for the purpose are given in detail as follows. The target is to map the state of many photons to a genuine 
single photon state as in Eq. (\ref{1}).

\subsection{Parity gate}
In Ref. \cite{parity} some of us propose an optical realization of unknown quantum state discrimination scheme in \cite{Bergou}. A crucial element there is a parity gate that sends the bi-photon components of different parity to the different paths. However, that gate actually selects out either even parity component (the linear combinations of $|HH\rangle$ and $|VV\rangle$) or odd parity component (the linear combinations of $|HV\rangle$ and $|VH\rangle$) through homodyne detection. We here present a correct design for such deterministic parity gate. On the other hand, parity gate is an important tool in quantum information processing. The wide usages of the gate with abstract quantum systems are discussed in \cite{Io}, and various optical realizations can be found in, for example, \cite{P-J-C, B-K, B-R, Nemoto, Munro, He}. We enrich the methods to realize parity gate with
a special feature that the different parity components will be obtained simultaneously on different paths.

\begin{figure}[ptb]
\includegraphics[width=6cm]{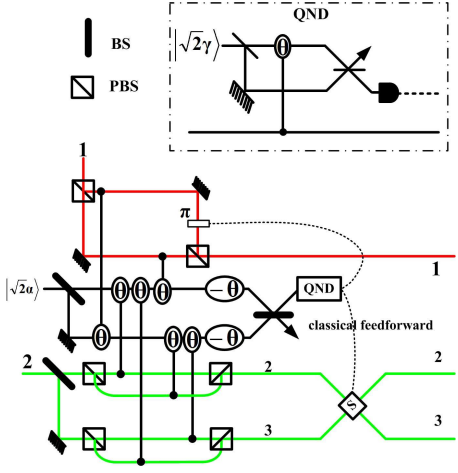}
\caption{(Color online) Parity gate. Here we use two qubus beams in coherent state $|\alpha\rangle$ to interact with the indicated photonic modes. Each coupling gives rise to a phase shift $\theta$. Two phase shifters $-\theta$ are applied to two qubus beams, respectively. The setup of the QND module is shown in dash-dotted line, where one of the beams $|\gamma\rangle$ is coupled to the first qubus beam after a 50:50 BS between two phase shifter $-\theta$ and the QND module. One output beam in the QND module is measured by a number-non-resolving detector. The phase shift $\pi$ on $|V\rangle_1$ and the switch of the path mode 2 and 3 are conditional on the classically feed-forwarded measurement results.}
\label{parity-p}
\end{figure}

The gate deals with an input state
\begin{align}
 \left\vert \psi\right\rangle _{in}=a\left\vert HH\right\rangle
_{12}+b\left\vert HV\right\rangle _{12}+c\left\vert VH\right\rangle
_{12}+d\left\vert VV\right\rangle _{12}
\label{in}
\end{align}
of two photons running on track 1 and 2.
We apply a 50:50 beam splitter (BS) to divide the second photon
into two spatial modes 2 and 3:
\begin{align}
\left\vert \psi\right\rangle _{in}&\rightarrow  \frac{1}{\sqrt{2}}a\left\vert H\right\rangle _{1}\left(  \left\vert
H\right\rangle _{2}+\left\vert H\right\rangle _{3}\right)  +\frac{1}{\sqrt{2}%
}b\left\vert H\right\rangle _{1}\left(  \left\vert V\right\rangle
_{2}+\left\vert V\right\rangle _{3}\right) \nonumber\\
&  +\frac{1}{\sqrt{2}}c\left\vert V\right\rangle _{1}\left(  \left\vert
H\right\rangle _{2}+\left\vert H\right\rangle _{3}\right)  +\frac{1}{\sqrt{2}%
}d\left\vert V\right\rangle _{1}\left(  \left\vert V\right\rangle
_{2}+\left\vert V\right\rangle _{3}\right).
\end{align}
Two qubus or communication beams $\left\vert \alpha\right\rangle \left\vert
\alpha\right\rangle $ are then introduced, and are coupled to the corresponding photonic modes through the cross-phase modulation (XPM) processes described by the Hamiltonian ${\cal H}=-\hbar\chi \hat{n}_i\hat{n}_j$, where $\chi$ is the nonlinear intensity and $\hat{n}_i$ ($\hat{n}_j$)
the number operator of the coupling modes.
The interaction pattern in Fig. \ref{parity-p} is summarized in Tab. \ref{tb1}.
\begin{table}
\begin{tabular}{|c|c|c|c|c|}\hline
 & $|H\rangle_2$ & $|V\rangle_2$ & $|H\rangle_3$ & $|V\rangle_3$\\ \hline
$|H\rangle_1$  & $\bigcirc$   &   &  &  $\bigcirc$   \\ \hline
$|V\rangle_1$  &  &$\bigtriangleup$  & $\bigtriangleup$ &   \\ \hline
\end{tabular}
\caption{$\bigcirc$ and $\bigtriangleup$ represent the coupling to the first and the second qubus beam, respectively. The XPM pattern in Fig. 1 is read as the first beam being coupled to $|H\rangle_1$ of the first photon and $\{|H\rangle_2, |V\rangle_3\}$ of the second photon, while the second beam to $|V\rangle_1$ of the first photon and $\{|V\rangle_2, |H\rangle_3\}$ of the second. }
\label{tb1}
\end{table}
Suppose the XPM phase shifts induced by the couplings are all $\theta$. A qubus beam will undergo a conditioned XPM phase shift $|\alpha\rangle\rightarrow |\alpha e^{i\theta}\rangle$ if coupled to one photonic mode, and $|\alpha\rangle\rightarrow |\alpha e^{2i\theta}\rangle$ if coupled to two modes. As the result, we will realize the following transformation of the total system:
\begin{align}
&~~~~~|\psi\rangle_{in}|\alpha\rangle|\alpha\rangle \nonumber\\
&\rightarrow  \frac{1}{\sqrt{2}}\left\vert H\right\rangle _{1}\left(  a\left\vert
H\right\rangle _{2}+b\left\vert V\right\rangle _{3}\right)  \left\vert \alpha
e^{2i\theta}\right\rangle \left\vert \alpha \right\rangle \nonumber\\
&+\frac{1}{\sqrt{2}}\left\vert V\right\rangle _{1}\left(  c\left\vert
H\right\rangle _{3}+d\left\vert V\right\rangle _{2}\right)  \left\vert \alpha
\right\rangle \left\vert \alpha e^{2i\theta}\right\rangle \nonumber\\
&  +\frac{1}{\sqrt{2}}\left\vert H\right\rangle _{1}\left(  a\left\vert
H\right\rangle _{3}+b\left\vert V\right\rangle _{2}\right)  \left\vert \alpha e^{i\theta}\right\rangle \left\vert \alpha e^{i\theta}\right\rangle
\nonumber\\
&  +\frac{1}{\sqrt{2}}\left\vert V\right\rangle _{1}\left(  c\left\vert
H\right\rangle _{2}+d\left\vert V\right\rangle _{3}\right)  \left\vert \alpha
e^{i\theta}\right\rangle \left\vert \alpha e^{i\theta}\right\rangle .
\end{align}

After that, a phase shifter of $-\theta$ is respectively applied to two qubus beams, and then one more 50:50 BS implements
the transformation $\left\vert
\alpha_{1}\right\rangle \left\vert \alpha_{2}\right\rangle \rightarrow
\left\vert \frac{\alpha_{1}-\alpha_{2}}{\sqrt{2}}\right\rangle \left\vert
\frac{\alpha_{1}+\alpha_{2}}{\sqrt{2}}\right\rangle $ of the coherent-state components.
The state of the total system will be therefore transformed to
\begin{align}
&~~~\frac{1}{\sqrt{2}}\left\vert H\right\rangle _{1}\left(  a\left\vert
H\right\rangle _{3}+b\left\vert V\right\rangle _{2}\right)  \left\vert 0\right\rangle \left\vert \sqrt{2}\alpha \right\rangle \nonumber\\
&+\frac{1}{\sqrt{2}}\left\vert V\right\rangle _{1}\left(  c\left\vert
H\right\rangle _{2}+d\left\vert V\right\rangle _{3}\right)  \left\vert 0\right\rangle \left\vert \sqrt{2}\alpha \right\rangle \nonumber\\
&  +\frac{1}{\sqrt{2}}\left\vert H\right\rangle _{1}\left(  a\left\vert
H\right\rangle _{2}+b\left\vert V\right\rangle _{3}\right)  \left\vert -\beta \right\rangle \left\vert \sqrt{2}\alpha \cos\theta\right\rangle
\nonumber\\
&  +\frac{1}{\sqrt{2}}\left\vert V\right\rangle _{1}\left(  c\left\vert
H\right\rangle _{3}+d\left\vert V\right\rangle _{2}\right)  \left\vert \beta\right\rangle \left\vert \sqrt{2}\alpha \cos\theta \right\rangle,
\label{output}
\end{align}
where $|\beta\rangle=|i\sqrt{2}\alpha\sin
\theta\rangle$.

The first coherent-state component in Eq. (\ref{output}) is either vacuum or a cat state (the superposition of $|\pm\beta\rangle$ in the second piece). The target output could be therefore obtained by the projection $\left\vert
n\right\rangle \left\langle n\right\vert $ on the first qubus beam.
If $n=0$, we will obtain
\begin{align}
 \left\vert \psi\right\rangle _{out}=a\left\vert HH\right\rangle
_{13}+b\left\vert HV\right\rangle _{12}+c\left\vert VH\right\rangle
_{12}+d\left\vert VV\right\rangle _{13},
\label{parity}
\end{align}
with the even parity component on path 1 and 3, and the odd parity component on path 1 and 2, respectively.
If $n\neq 0$, on the other hand, there will be the output
\begin{align}
\left\vert \psi\right\rangle _{out}&=e^{-in\frac{\pi}{2}
}\{\left\vert H\right\rangle _{1}\left(  a\left\vert H\right\rangle
_{2}+b\left\vert V\right\rangle _{3}\right)\nonumber\\
&+e^{in\pi}\left\vert
V\right\rangle _{1}\left(  c\left\vert H\right\rangle _{3}+d\left\vert
V\right\rangle _{2}\right)\},
\label{parity1}
\end{align}
which can be transformed to the form in Eq. (\ref{parity}) by a phase shift $\pi$ on $|V\rangle_1$ following the classically feed-forwarded measurement
result $n$ and a switch of the path modes 2 and 3.

Photon number resolving detection on one of the output qubus beams is crucial in realizing the states in Eqs. (\ref{parity}) and (\ref{parity1}).
The general aspect and development of photon number resolving detection can be found in \cite{P}. The best device thus far to simulate the projector $\left\vert n\right\rangle \left\langle n\right\vert $ is transition edge sensor (TES), a superconducting microbolometer that has demonstrated very high detection efficiency (95\% at $\lambda=1550$ nm) and high photon number resolution\cite{detector, d2,d3,d4}. To realize the projection $\left\vert n\right\rangle \left\langle n\right\vert $ deterministically, we could apply the indirect measurement method in \cite{He}. It is to use a quantum non-demolition detection (QND) module shown inside dash-dotted line in Fig. \ref{parity-p}.
In the module one of the coherent state $|\gamma\rangle$ is coupled to the first coherent-state component in Eq. (\ref{output}), making the
XPM phase shifts
\begin{equation}
\left\vert \pm\beta\right\rangle \left\vert \gamma\right\rangle \left\vert
\gamma\right\rangle \rightarrow e^{-\left\vert \beta\right\vert ^{2}%
/2}\overset{\infty}{\underset{n=0}{{\sum}}}\frac{\left(  \pm\beta\right)
^{n}}{\sqrt{n!}}\left\vert n\right\rangle \left\vert \gamma e^{in\theta
}\right\rangle \left\vert \gamma\right\rangle
\label{pro}
\end{equation}
conditioned on the Fock states $|n\rangle$ in the first qubus beam of Eq. (\ref{output}).
A 50:50 BS then maps the coherent-state components on the right hand side of Eq. (\ref{pro}) to $\left\vert
\frac{\gamma e^{in\theta}-\gamma}{\sqrt{2}}\right\rangle \left\vert
\frac{\gamma e^{in\theta}+\gamma}{\sqrt{2}}\right\rangle $ ($n=0,1,\cdots
,\infty$). We will detect $n$ by using a photon number non-resolving detector described by the POVM
elements \cite{n}
\begin{align}
\Pi_{0}  &  =\overset{\infty}{\underset{n=0}{{\sum}}}\left(  1-\eta\right)
^{n}\left\vert n\right\rangle \left\langle n\right\vert ,\nonumber\\
\Pi_{1}  &  =I-\Pi_{0},
\label{povm}
\end{align}
where $\eta<1$ is the quantum efficiency of the detector, and $\Pi_{0}$ and $\Pi_{1}$ correspond to detecting no photon and any number of photon, respectively.
Since each of the states $\left\vert \frac{\gamma
e^{in\theta}-\gamma}{\sqrt{2}}\right\rangle $ has a certain distribution of photon numbers (Poisson peak), the action of $\Pi_1$ on it will
be actually the operator, $\Pi_{1,k}=\sum_{m=n_k}^{n_k'}(1-(1-\eta)^m)|m\rangle\langle m|$, if the dominant distribution for the peak $n=k$ is from
$n_k$ to $n_k'$. We let the detector respond to a $\left\vert \frac{\gamma e^{ik\theta}-\gamma}{\sqrt{2}}\right\rangle $ according to the total photon detection probability $\langle \frac{\gamma e^{ik\theta}-\gamma}{\sqrt{2}}|\Pi_{1,k}|\frac{\gamma e^{ik\theta}-\gamma}{\sqrt{2}}\rangle$. This response, as a function of the photon detection probability, can be a measured quantity (for example, voltage or current converted from the measured light) by the detector. Such photon number non-resolving detector could read the sufficiently large difference in the intensities of the measured coherent light, though it is unable to distinguish between pulses of close photon numbers (for example, two Fock states $|n\rangle$ and $|n+m\rangle$ if $m$ is not large enough). We can use a sufficiently large $|\gamma|$ such that the photon number distributions of $\left\vert \frac{\gamma e^{in\theta}-\gamma}{\sqrt{2}}\right\rangle $ will be fairly separated and the reactions of the detector to them will be mutually distinct. As the result, the operator $\Pi_{1,k}$ will indirectly project out the Fock state $|k\rangle$ in the first qubus beam by a particular response of the detector to $\left\vert \frac{\gamma
e^{ik\theta}-\gamma}{\sqrt{2}}\right\rangle $.

\begin{figure}[ptb]
\includegraphics[width=6cm]{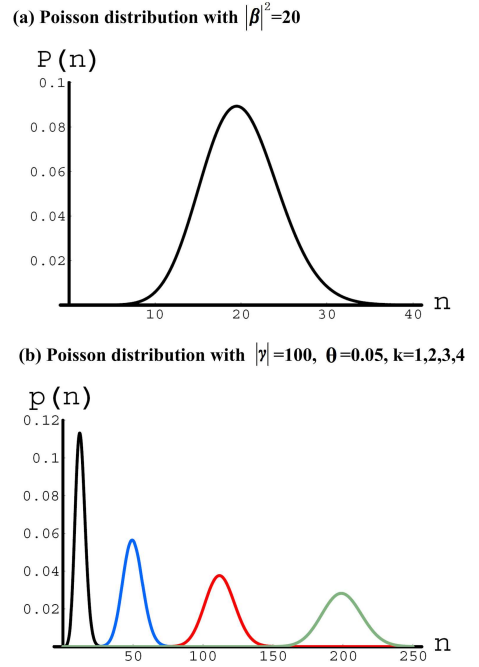}
\caption{(Color online) Example of the photon number resolving measurement on an output qubus beam with the average photon number $|\beta|^2=20$. (a) Poisson distribution of the measured output qubus beam photon numbers. Its dominant photon numbers are between $n=8$ and $n=35$. These photon numbers occurring with the shown probabilities give rise to so many possibly measured quantities by a number non-resolving detector described by Eq. (\ref{povm}). So there should be $27$ classically feed-forwarded measurement results in Fig. \ref{parity-p}. (b) The Poisson peaks of $\left\vert \frac{\gamma e^{ik\theta}-\gamma}{\sqrt{2}}\right\rangle $ with $k=1,2,3,4$. A qubus beam with the amplitude $|\gamma|=100$ in the QND module and an XPM phase shift $\theta=0.05$ in Eq. (\ref{pro}) will be sufficient to have the negligible overlaps for their photon number distributions. }
\label{distribution}
\end{figure}

In Fig. \ref{distribution} we show such an example of the separate photon number distributions induced by different $|k\rangle$ ($k=1,2,\cdots$) in the first qubus beam when $\left\vert\gamma\right\vert =100$ and $\theta=0.05$. Also from the figure the number of possibly measured results by a detector is finite, as the photon number Poisson distribution of $|\pm \beta\rangle$ is within a finite range. Once the possibly measured quantities are set in design,
we could use them to check by the deviations if the gate works correctly.
The error probability of this parity gate is
\begin{align}
P_E&=||\sum_{n=0}^{\infty}e^{-\left\vert \beta\right\vert ^{2}%
/2}\frac{\left(  \pm\beta\right)
^{n}}{\sqrt{n!}}\left\vert n\right\rangle \Pi_{0}^{\frac{1}{2}} \left\vert \frac{\gamma
e^{in\theta}-\gamma}{\sqrt{2}}\right\rangle||^2 \nonumber\\
&\sim exp\{-2(1-e^{-\frac{1}{2}\eta\gamma^2 \theta^2})\alpha^2\sin^2\theta\},
\end{align}
considering $\theta \ll 1$ from a weak cross-Kerr nonlinearity, so it could deterministically send the different parity components of a bi-photon state to different paths given that $2\alpha^2\sin^2\theta \gg 1$ and $\frac{1}{2}\eta \gamma^2\theta^2\gg 1$.

Other methods to perform the projection $|n\rangle\langle n|$ indirectly with QND module include quadrature $\hat{x}$ ($\hat{p}$) measurement on $\left\vert\frac{\gamma e^{in\theta}-\gamma}{\sqrt{2}}\right\rangle$ or phase measurement on $\left\vert \gamma e^{in\theta
}\right\rangle$ via homodyne-heterdyne techniques \cite{Nemoto}. For a deterministic operation, the condition $\beta \theta^2\gg 1$ in \cite{Nemoto}, which is from $\hat{x}$ measurement on the first qubus beam directly, can be relaxed if the beams of a sufficiently large $|\gamma|$ are used in the QND module.

If we project out the even or odd parity component of the output bi-photon state (on only one pair of tracks 1, 2 or 1, 3) by a QND device, the gate will also work as an entangler for separable input bi-photon states. The main advantage of using two qubus beams as in \cite{He} is that an XPM phase shift of $-\theta$ (which is only possible with a coupling constant $\chi$ of the opposite sign) or the displacement operation on qubus beam in some other qubus mediated parity gates \cite{Nemoto, Munro} can be avoided. With EIT techniques \cite{EIT}, it is possible to use a realistic qubus beam intensity and maintain a good coherence of the total quantum state against the losses in generating a small $\theta$ through such XPM processes \cite{loss}. Another advantage of the design is that qubus beams can be recycled. Since the $\theta$ induced by weak cross-Kerr nonlinearity is small, i.e., $\left\vert \sqrt{2}\alpha\cos
\theta\right\vert \simeq\left\vert \sqrt{2}\alpha\right\vert $, the other
unmeasured qubus beam can be used again until it is consumed too much after many
times of detection.

\subsection{Variation of parity gate---controlled-path gate}

\begin{figure}[ptb]
\includegraphics[width=8cm]{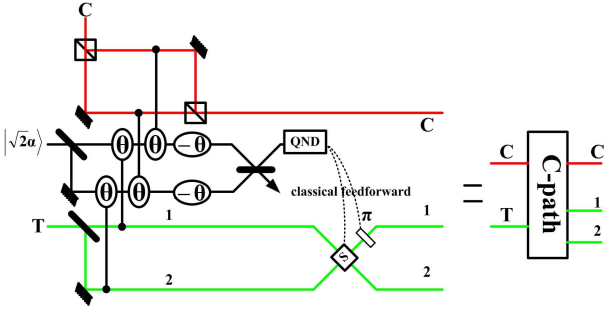}
\caption{(Color online) Controlled-path gate. The two qubus beams interact with the indicated photonic modes through cross-Kerr nonlinearity creating $\theta$. A phase shifters $-\theta$ is respectively applied to two qubus beams. The QND module performs the photon number resilving detection, the results of which are classically feed-forwarded to control the switch of path 1 and 2 and a phase shift $\pi$ on path 1. The effect of the gate is realizing the control on the paths of the target photon $T$ by the polarizations of the control photon $C$. }
\label{c-path-p}
\end{figure}

We will slightly modify the parity gate to have a deterministic realization of Controlled-path gate (C-path gate) introduced in \cite{Lin}.
This gates implements the transformation ($C$ stands for the control and $T$ the target in Fig. \ref{c-path-p}):
\begin{align}
&  \left\vert \psi\right\rangle _{CT}=a\left\vert HH\right\rangle
_{CT}+b\left\vert HV\right\rangle _{CT}+c\left\vert VH\right\rangle
_{CT}+d\left\vert VV\right\rangle _{CT}\nonumber\\
&  \rightarrow a\left\vert HH\right\rangle _{C1}+b\left\vert HV\right\rangle
_{C1}+c\left\vert VH\right\rangle _{C2}+d\left\vert VV\right\rangle
_{C2}=\left\vert \phi\right\rangle ,
\label{c-path}
\end{align}
where the index $1$ and $2$ denotes two different paths, controlling the paths of the target single-photon qubit $T$ by the polarizations of the control photon $C$.

The process is similar to that in parity gate. We first use a 50:50 beam splitter
(BS) to divide the target photon into two spatial modes 1 and 2. Then two
qubus beams $\left\vert \alpha\right\rangle \left\vert \alpha\right\rangle $
are used to interact with the corresponding photonic modes in the pattern summarized in Tab. \ref{tb2}.
\begin{table}
\begin{tabular}{|c|c|c|}\hline
 & $|H/V\rangle_1$ & $|H/V\rangle_2$ \\ \hline
$|H\rangle_C$  &    &    $\bigtriangleup$    \\ \hline
$|V\rangle_C$  & $\bigcirc$  &  \\ \hline
\end{tabular}
\caption{$\bigcirc$ and $\bigtriangleup$ represent the coupling to the first and the second qubus beam, respectively. The XPM pattern in C-path gate is that the first qubus beam is coupled to $|V\rangle_C$ of the control photon and all modes of the target photon on path 1 after BS, while the second beam to $|H\rangle_C$ of the control photon and those on path 2 from the target photon. }
\label{tb2}
\end{table}
The phase shifters of  $-\theta$  follow up to change the phases of the coherent-state components to the following
\begin{align}
&  \frac{1}{\sqrt{2}}\left\vert \phi\right\rangle \left\vert \alpha
\right\rangle \left\vert \alpha\right\rangle
+\frac{1}{\sqrt{2}}\left\vert H\right\rangle _{C}\left(  a\left\vert
H\right\rangle _{2}+b\left\vert V\right\rangle _{2}\right)  \left\vert \alpha
e^{-i\theta}\right\rangle \left\vert \alpha e^{i\theta}\right\rangle
\nonumber\\
&  +\frac{1}{\sqrt{2}}\left\vert V\right\rangle _{C}\left(  c\left\vert
H\right\rangle _{1}+d\left\vert V\right\rangle _{1}\right)  \left\vert \alpha
e^{i\theta}\right\rangle \left\vert \alpha e^{-i\theta}\right\rangle .
\end{align}
Finally, one more 50:50 BS converts the coherent-state components to
\begin{align}
& \frac{1}{\sqrt{2}}\{\vert H \rangle _{C}(a \vert
H \rangle _{2}+b \vert V \rangle _{2})  \vert-\beta\rangle \vert \sqrt{2}\alpha\cos\theta\rangle \nonumber\\
& +\vert V \rangle _{C}(c\vert
H\rangle _{1}+d \vert V\rangle _{1})
\vert\beta \rangle \vert \sqrt{2}\alpha\cos\theta\rangle\} \nonumber\\
&+\frac{1}{\sqrt{2}} \vert \phi\rangle \vert 0\rangle \vert \sqrt
{2}\alpha\rangle ,
\label{2}
\end{align}
where $\left\vert \beta\right\rangle =\left\vert i\sqrt{2}\alpha\sin
\theta\right\rangle $. Then, we could use the QND module in Fig. \ref{parity-p} to perform the projection $|n\rangle\langle n|$
for achieving the desired output state $\left\vert \phi\right\rangle $ deterministically.

As we will discuss in what follows, this C-path gate works as a building block for constructing various logic gates.
Next, we will apply it in a teleportation scheme to transform a multi-photon state to the corresponding single photon state.

\subsection{Teleportation scheme for unknown state discrimination}
\label{tele}

Now, we present an improved teleportation scheme on that in \cite{parity}, which transforms a multi-photon state to the corresponding single photon state. We illustrate the procedure first with the simplest example of a two-photon input state, $|\psi_1\rangle|\psi_2\rangle$, where $|\psi_i\rangle=c_i|H\rangle+d_i|V\rangle$, and the coefficients $c_i$ and $d_i$ could be unknown. Using the same ancilla resources,
a single photon state $\frac{1}{\sqrt{2}}(|H\rangle_C+|V\rangle_C)$ and a photon pair in Bell state $|\Phi^{+}\rangle_{12}=\frac{1}{\sqrt{2}}(|HH\rangle_{12}+|VV\rangle_{12})$, as in \cite{parity}, we process them with a first C-path gate as follows:
\begin{align}
&\frac{1}{\sqrt{2}}(|H\rangle_C+|V\rangle_C)\frac{1}{\sqrt{2}}(|HH\rangle_{12}+|VV\rangle_{12})\nonumber\\
&\rightarrow \frac{1}{2}( |H, H, H\rangle_{C32}+|H, V, V\rangle_{C32}
+|V, H, H\rangle_{C42}\nonumber\\
&+|V, V, V\rangle_{C42})=|\Sigma\rangle,
\end{align}
where the $|H\rangle_C$ and $|V\rangle_C$ of the control single photon control the first photon in the Bell pair to different paths 3 and 4.
The state $|\Sigma\rangle$ can be rewritten as
\begin{align}
&\frac{1}{\sqrt{2}}(|H\rangle_C |0\rangle_S+|V\rangle_C |1\rangle_S)\frac{1}{\sqrt{2}}(|HH\rangle_{12}+|VV\rangle_{12})\nonumber\\
&=\frac{1}{2}( |H, H, H\rangle_{C32}+|H, V, V\rangle_{C32}
+|V, H, H\rangle_{C42}\nonumber\\
&+|V, V, V\rangle_{C42})
\label{bell},
\end{align}
a product of two Bell states $|\Phi^{+}\rangle_{CS}|\Phi^{+}\rangle_{12}$. In the first effective Bell state we adopt the notation in \cite{parity} to define the switch state $|0\rangle_S$,
which sends the modes on path 1 to path 3, and $|1\rangle_S$, which sends the modes on path 1 to path 4 at the same time.
The effect of the C-path gate is to create an effective Bell state $|\Phi^{+}\rangle_{CS}$.
Suppose two single photons in the state $|\psi_i\rangle$ are placed on track A and B, respectively. Then we will have the following total state \cite{notation}:
\begin{align}
&|\psi_1\rangle_A|\psi_2\rangle_B |\Sigma\rangle \nonumber\\
&= \frac{1}{4}(|\Phi^+\rangle_{AC}|\psi_1\rangle_S+|\Psi^+\rangle_{AC}~\sigma_x|\psi_1\rangle_S+|\Psi^-\rangle_{AC}~(-i\sigma_y)|\psi_1\rangle_S\nonumber\\
&+|\Phi^-\rangle_{AC}~\sigma_z|\psi_1\rangle_S )(|\Phi^+\rangle_{B2}|\psi_2\rangle_1+|\Psi^+\rangle_{B2}~\sigma_x|\psi_2\rangle_1\nonumber\\
&+|\Psi^-\rangle_{B2}~(-i\sigma_y)|\psi_2\rangle_1
+|\Phi^-\rangle_{B2}~\sigma_z|\psi_2\rangle_1 ).
\end{align}
Performing the Bell-state measurements on the modes $\{A,C\}$ and $\{B,2\}$ and classically-feedforwarding the results for the post operations,
we will finally obtain the state
\begin{align}
&(c_1|0\rangle_S+d_1|1\rangle_S)(c_2|H\rangle_1+d_2|V\rangle_1)\nonumber\\
&=c_1c_2|H\rangle_3+c_1d_2|V\rangle_3+d_1c_2|H\rangle_4+d_1d_2|V\rangle_4,
\end{align}
which is a single photon qudit inheriting the coefficients of the input bi-photon state $|\psi_1\rangle|\psi_2\rangle$.

\begin{figure}[ptb]
\includegraphics[width=7cm]{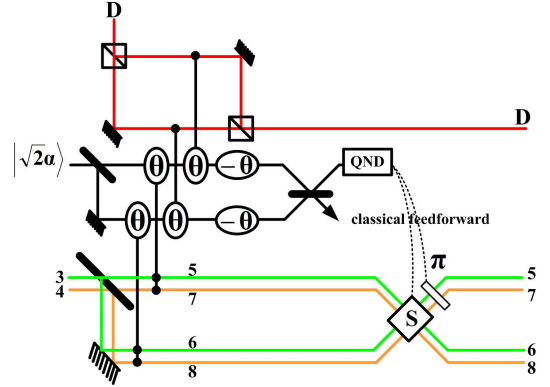}
\caption{(Color online) A modified C-path gate from that in Fig. \ref{c-path-p}. Here the polarizations of photon $D$ simultaneously control the path 3 and path 4 mode of the first photon of an entangled pair. Then the modes on path 3 and 4 will be converted to the modes on path 5, 6, 7 and 8. }
\label{c-path-2}
\end{figure}

To teleport a triple-photon state $|\psi_1\rangle|\psi_2\rangle|\psi_3\rangle$ to the corresponding single photon qudit, we will use an extra ancilla photon in the state $\frac{1}{\sqrt{2}}(|H\rangle_D+|V\rangle_D)$. We let it control the paths of the first photon of the Bell pair, which runs on path 3 and 4 after the first C-path gate.
This control should be performed on path 3 and 4 together, so we modify the standard C-path gate to a C-path-2 gate shown in
Fig. \ref{c-path-2}.
In this gate a 50:50 BS splits the modes on path 3 to path 5 and 6, while it devides the modes on path 4 to path 7 and 8 (the global coefficient is neglected):
\begin{align}
&~~~~~(|H\rangle_D+|V\rangle_D)|\Sigma\rangle \nonumber\\
&\rightarrow (|H\rangle_D+|V\rangle_D)(|HH\rangle_{C2}\frac{1}{\sqrt{2}}(|H\rangle_5+|H\rangle_6)\nonumber\\
&+|HV\rangle_{C2}\frac{1}{\sqrt{2}}(|V\rangle_5+|V\rangle_6)
+|VH\rangle_{C2}\frac{1}{\sqrt{2}}(|H\rangle_7+|H\rangle_8)\nonumber\\
&+|VV\rangle_{C2}\frac{1}{\sqrt{2}}(|V\rangle_7+|V\rangle_8)).
\end{align}
The interaction pattern for the qubus beams with the above state is given in Tab. \ref{tb3}.
\begin{table}[ptb]%
\begin{tabular}
[c]{|c|c|c|c|c|}\hline
& $|H/V\rangle_{5}$ & $|H/V\rangle_{6}$& $|H/V\rangle_{7}$ & $|H/V\rangle_{8}$\\\hline
$|H\rangle_{D}$ &  & $\bigtriangleup$ & & $\bigtriangleup$\\\hline
$|V\rangle_{D}$ & $\bigcirc$ &  & $\bigcirc$ &\\\hline
\end{tabular}
\caption{$\bigcirc$ and $\bigtriangleup$ represent the coupling to the first
and the second qubus beam, respectively. The XPM pattern in C-path-2 gate is
that the first qubus beam is coupled to $|V\rangle_{D}$ of the first photon
and all modes of the third photon on path $5$ and $7$ after
BS, while the second beam to $|H\rangle_{D}$ of the first photon and those on
path $6, 8$ from the third photon. }%
\label{tb3}%
\end{table}
Through the similar post-selection to that in a standard C-path gate, the ancilla photon on track D will control the paths of the first photon of the Bell pair to path 5, 6, 7 and 8 with its polarization mode $|H/V\rangle_D$. This process is equivalent to adding another effective Bell state as in the first C-path gate.
After the corresponding operations according to the Bell state measurements, any POVM can be performed by linear optical circuits on the single photon running on path 5, 6, 7 and 8, which inherits the structure of the input $|\psi_1\rangle|\psi_2\rangle|\psi_3\rangle$ \cite{h-b-w, parity}. The C-path gate teleportation method thus greatly simplfies the optical realization of unknown qubit discrimination in \cite{parity}.

\section{Purely circuit-based approach}
\label{sec3}

In teleporting an n-photon state to the corresponding single photon state, $n+1$ ancilla photons are necessary as in the above-discussed scheme for unknown state discrimination. These ancilla resources can be saved by the approach discussed in this section. Here we illustrate the method with the examples of transforming two-photon and triple-photon states to the corresponding single photon states, which can be processed by linear optics.

\subsection{Two-photon state}

\begin{figure}[ptb]
\includegraphics[width=7cm]{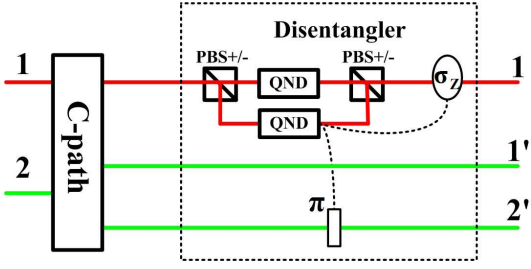}
\caption{(Color online) Setup to transform a double photon state in the form of Eq. (\ref{in}) to that in Eq. (\ref{2-trans}). After a C-path gate, the modes $|\pm\rangle$ of the first photon are detected by the QND modules. The measurement results are forwarded to transform the output to the form in Eq. (\ref{2-trans}) by an operation $\sigma_z$ on the first photon and a phase shift $\pi$ on the second. The part in dashed line is called Disentangler because it transforms arbitary input two-photon state to a tensor product.}
\label{two-photon}
\end{figure}

The schematic setup for two-photon transformation is shown in Fig. \ref{two-photon}. After a C-path gate, a two-photon input
state $\left\vert\psi\right\rangle _{12}$ in the form of that in Eq. (\ref{in}) is transformed to
\begin{align}
\left\vert \phi\right\rangle &=  \frac{1}{\sqrt{2}}\left\vert +\right\rangle _{1}\left(  a\left\vert
H\right\rangle _{1^{^{\prime}}}+b\left\vert V\right\rangle _{1^{^{\prime}}%
}+c\left\vert H\right\rangle _{2^{^{\prime}}}+d\left\vert V\right\rangle
_{2^{^{\prime}}}\right) \nonumber\\
&  +\frac{1}{\sqrt{2}}\left\vert -\right\rangle _{1}\left(  a\left\vert
H\right\rangle _{1^{^{\prime}}}+b\left\vert V\right\rangle _{1^{^{\prime}}%
}-c\left\vert H\right\rangle _{2^{^{\prime}}}-d\left\vert V\right\rangle
_{2^{^{\prime}}}\right) ,
\end{align}
with the first photon expressed in the basis $|\pm\rangle=\frac{1}{\sqrt{2}}(|H\rangle\pm |V\rangle)$.
Then we process it with the circuit part called Disentangler shown in dotted line in Fig. \ref{two-photon}.
Disentangler includes a balanced Mach-Zehnder (MZ)
interferometer formed with two polarizing beam splitters (PBS$_{\pm}$), which transmit
$\left\vert +\right\rangle $ and reflect
$\left\vert -\right\rangle $, and two QND modules in each arms, respectively.
If the component $|+\rangle$ is detected by a QND module,
the following state,
\begin{equation}
\left\vert +\right\rangle _{1}\left(  a\left\vert H\right\rangle
_{1^{^{\prime}}}+b\left\vert V\right\rangle _{1^{^{\prime}}}+c\left\vert
H\right\rangle _{2^{^{\prime}}}+d\left\vert V\right\rangle _{2^{^{\prime}}
}\right),
\label{2-trans}
\end{equation}
will be projected out.
On the other hand, if $|-\rangle$ is detected, a phase shift $\pi$ will be performed on path $2^{\prime}$
of the second photon to get the same final state. Now, the state of the second photon inherits the coefficients of the initial two-photon state $\left\vert \psi\right\rangle _{12}$. Since we apply the detections with QND modules, the other photon is also preserved as $|+\rangle_1$ in Eq. (\ref{2-trans}).

\subsection{Triple-photon and multi-photon state}
\label{m1}

\begin{figure}[ptb]
\includegraphics[width=7.8cm]{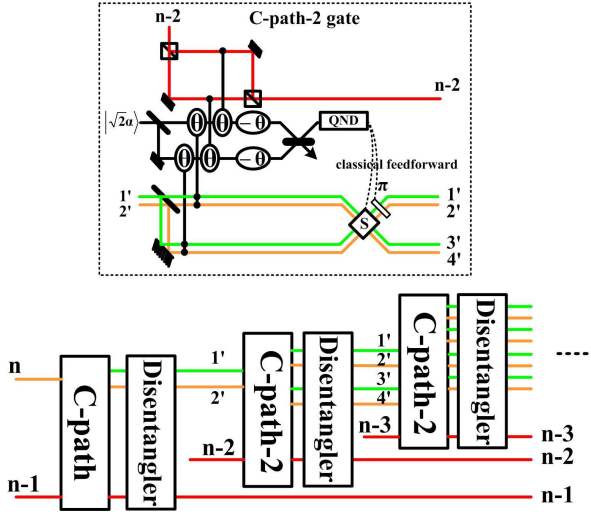}
\caption{(Color online) Schematic setup for transforming a product of $n$ single photon qubits to that in Eq. (\ref{m-out}). The first C-path-2 gate from the left is illustrated in dotted line, and the coupling pattern of the qubus beams in all other C-path-2 gates is similar. The Disentanglers are similar to that in Fig. \ref{two-photon}. The polarizations of the $(n-1)$-th photon controls the paths of the $n$-th photon through a C-path gate, and the polarizations of the $(n-2)$-th photon continues to control the $n$-th photon to doubled number of paths through a C-path-2 gate, and so forth to the control by the first photon. Through such iterative operations, the output state of the $n$-th photon will inherit the coefficients of the input multi-photon state. The total number of C-path gates (including C-path-2), as well as that of the Disentanglers, after $n-1$ times of control is $n-1$.}
\label{multi-photon}
\end{figure}

This method can be generalized to multi-photon state as shown in Fig. \ref{multi-photon}.
For simplicity, we only give the detail of the triple-photon transformation.
The input triple-photon state is given as
\begin{equation}
\left\vert \psi\right\rangle _{123}=\left\vert HH\right\rangle \left\vert
\phi_{1}\right\rangle +\left\vert HV\right\rangle \left\vert \phi
_{2}\right\rangle +\left\vert VH\right\rangle \left\vert \phi_{3}\right\rangle
+\left\vert VV\right\rangle \left\vert \phi_{4}\right\rangle ,
\end{equation}
where $\left\vert \phi_{j}\right\rangle =\alpha_{j}\left\vert H\right\rangle
+\beta_{j}\left\vert V\right\rangle $ $\left(  j=1,2,3,4\right)$, and $\sum_{j=1}^4|\alpha_j|^2+\sum_{j=1}^4|\beta_j|^2=1$.
Firstly, similar to the case of two-photon state, we implement a C-path gate
and a Disentangler to the second and third photon, to project out the following
state,%
\begin{equation}
\left\vert +\right\rangle _{2}\left[  \left\vert H\right\rangle _{1}\left(
\left\vert \phi_{1}\right\rangle _{1^{^{\prime}}}+\left\vert \phi
_{2}\right\rangle _{2^{^{\prime}}}\right)  +\left\vert V\right\rangle
_{1}\left(  \left\vert \phi_{3}\right\rangle _{1^{^{\prime}}}+\left\vert
\phi_{4}\right\rangle _{2^{^{\prime}}}\right)  \right]  .
\end{equation}
Second, we use a C-path-2 gate in Fig. \ref{multi-photon} for further processing. The XPM pattern in this C-path-2 gate is
that the first qubus beam is coupled to $|V\rangle_{1}$ of the first photon
and all modes of the third photon on path $1^{^{\prime}},2^{^{\prime}}$ after
BS, while the second beam to $|H\rangle_{1}$ of the first photon and those on
path $3^{^{\prime}},4^{^{\prime}}$ from the third photon.
In the C-path-2 gate, one 50:50 BS implements the maps $1^{^{\prime}}\rightarrow1^{^{\prime}},3^{^{\prime}}$ and
$2^{^{\prime}}\rightarrow2^{^{\prime}},4^{^{\prime}}$ of the two spatial modes
of the third photon. Then two qubus beams $\left\vert \alpha\right\rangle
\left\vert \alpha\right\rangle $ are used to interact with the corresponding
photonic modes as indicated in Fig. \ref{multi-photon}.
The phase shifters of $-\theta$ continue to shift the phases of
the coherent-state components to those in the following:
\begin{align}
&  \frac{1}{\sqrt{2}}\left\vert +\right\rangle _{2}\left[  \left\vert
H\right\rangle _{1}\left(  \left\vert \phi_{1}\right\rangle _{1^{^{\prime}}%
}+\left\vert \phi_{2}\right\rangle _{2^{^{\prime}}}\right)  \left\vert
\alpha\right\rangle \left\vert \alpha\right\rangle \right.  \nonumber\\
&  +\left\vert V\right\rangle _{1}\left(  \left\vert \phi_{3}\right\rangle
_{3^{^{\prime}}}+\left\vert \phi_{4}\right\rangle _{4^{^{\prime}}}\right)
\left\vert \alpha\right\rangle \left\vert \alpha\right\rangle \nonumber\\
&  +\left\vert H\right\rangle _{1}\left(  \left\vert \phi_{1}\right\rangle
_{3^{^{\prime}}}+\left\vert \phi_{2}\right\rangle _{4^{^{\prime}}}\right)
\left\vert \alpha e^{-i\theta}\right\rangle \left\vert \alpha e^{i\theta
}\right\rangle \nonumber\\
&  \left.  +\left\vert V\right\rangle _{1}\left(  \left\vert \phi
_{3}\right\rangle _{1^{^{\prime}}}+\left\vert \phi_{4}\right\rangle
_{2^{^{\prime}}}\right)  \left\vert \alpha e^{i\theta}\right\rangle \left\vert
\alpha e^{-i\theta}\right\rangle \right]  .
\end{align}
Next, a 50:50 BS transforms two qubus beams, and a QND module performs $|n\rangle\langle n|$ on one the beams after that.
The following state,
\begin{equation}
\left\vert +\right\rangle _{2}\left[  \left\vert H\right\rangle _{1}\left(
\left\vert \phi_{1}\right\rangle _{1^{^{\prime}}}+\left\vert \phi
_{2}\right\rangle _{2^{^{\prime}}}\right)  +\left\vert V\right\rangle
_{1}\left(  \left\vert \phi_{3}\right\rangle _{3^{^{\prime}}}+\left\vert
\phi_{4}\right\rangle _{4^{^{\prime}}}\right)  \right],
\end{equation}
will be thus obtained with a conditioned switch and a phase shift $\pi$ on the modes $1^{\prime}$ and $2^{\prime}$ together accroding to the measured $n$ by the QND module. Like in the process in two-photon transformation, we will then use Disentangler on the first and third photon to realize the following state
\begin{equation}
\left\vert +\right\rangle _{1}\left\vert +\right\rangle _{2}\left(  \left\vert
\phi_{1}\right\rangle _{1^{^{\prime}}}+\left\vert \phi_{2}\right\rangle
_{2^{^{\prime}}}+\left\vert \phi_{3}\right\rangle _{3^{^{\prime}}}+\left\vert
\phi_{4}\right\rangle _{4^{^{\prime}}}\right)  .
\label{triple}
\end{equation}
Now, the third photon is in the desired single photon state carrying multiple spatial
modes, and it inherits the information from the polarizations of two other photons.

It is straightforward to generalize to the transformation
for a general multi-photon state
\begin{equation}
\left\vert \psi\right\rangle _{12\cdots n}=\left(  \left\vert H\cdots
H\right\rangle \left\vert \phi_{1}\right\rangle +\cdots+\left\vert V\cdots
V\right\rangle \left\vert \phi_{2^{n-1}}\right\rangle \right)  _{12\cdots n},
\end{equation}
in which the n-th photon is the superposition of $2^{n-1}$ pieces.
The process starts with a triple-photon transformation
applied to the (n-2)-th, the (n-1)-th and the n-th photons, and repeatedly apply the
C-path-2 gates to the other photons and the n-th photon, with all spatial
modes of the n-th photon split by 50:50 BS in each C-path-2
gates. The target state,
\begin{equation}
\left\vert +\right\rangle _{1}\otimes\cdots\otimes\left\vert +\right\rangle
_{n-1}\left(  \left\vert \phi_{1}\right\rangle _{1^{\prime}}+\cdots+\left\vert
\phi_{2^{n-1}}\right\rangle _{2^{n-1\prime}}\right)  ,
\label{m-out}
\end{equation}
will be obtained, with the path modes of the n-th photon inheriting the coefficients from other photons' modes.

In this transformation for $n$-photon state, $n-1$ C-path gates (one standard C-path gate, $n-2$
C-path-2 gates) and $n-1$ Disentangles are used, so the circuit resources are linearly dependent on the photon number of the input state.
Another significant advantage of the transformation is that no ancilla is necessary. Such transformation can be applied to realize unambiguous discrimination of multi-photon states including the unknown state discrimination schemes
in \cite{Bergou, h-b,h-b07}.

\section{Application in quantum network construction}
\label{sec4}

One important application of the transformation methods discussed in the previous sections is the realization of quantum logic gates. In what follows, we will illustrate how to construct various logic gates that are key to quantum computation.

\subsection{Merging gate}

\begin{figure}[ptb]
\includegraphics[width=8cm]{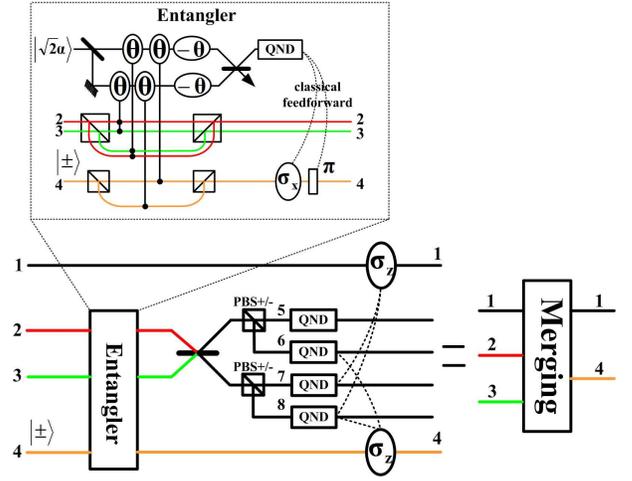}
\caption{(Color online) Merging gate. Two qubus beams interact with the modes of a single photon on path 2, 3 and those of the ancilla photon on path 4 as indicated in the part of Entangler. After Entangler, a 50:50 BS implements the interference of the single phootn modes on path 2, 3. Then the detection results of the QND modules will revise the possible output states in Eq. (\ref{merge-out}) to the standard form. The effect of the gate is realizing the merge of the single photon modes on path 2 and 3 to the same path. The gate also outputs a single photon in the state $|\pm\rangle$ after the QND detections, and this photon can be used in the next Merging gate.}
\label{merge}
\end{figure}

In the realization of logic gates, we should apply another building block---Merging gate.
The qubits we use are encoded as the superposition of $|H\rangle$ and $|V\rangle$, and they will be in extra spatial modes after a C-path gate.
A Merging gate can be be viewed as performing the inverse process of a C-path gate to merge a single photon in multiple patial modes back to only one spatial mode without changing anything else. Specifically, the schematic setup to deterministically realize the gate in Fig. \ref{merge} implements the transformation
\begin{align}
\left\vert \psi\right\rangle_{in}  &  =a\left\vert HH\right\rangle _{12}%
+b\left\vert HV\right\rangle _{12}+c\left\vert VH\right\rangle _{13}%
+d\left\vert VV\right\rangle _{13}\nonumber\\
&  \rightarrow a\left\vert HH\right\rangle _{14}+b\left\vert HV\right\rangle
_{14}+c\left\vert VH\right\rangle _{14}+d\left\vert VV\right\rangle _{14},
\end{align}
i.e., the merging of the second photon modes on path 2 and 3 to path 4. Here an extra single
photon in the state $\left\vert \pm\right\rangle =\frac{1}{\sqrt{2}}\left(
\left\vert H\right\rangle \pm\left\vert V\right\rangle \right)  $ should be used.

Suppose the initial state is $\left\vert \psi\right\rangle_{in} \left\vert +\right\rangle_4 $.
It is first processed by the part called Entangler in Fig \ref{merge}.
There two input photons will interact with two qubus beams $\left\vert \alpha\right\rangle \left\vert \alpha\right\rangle
$ as shown in the dashed line of Fig. \ref{merge}). The interaction pattern is summarized in Tab. \ref{tb4}. \begin{table}[ptb]%
\begin{tabular}
[c]{|c|c|c|c|c|}\hline
& $|H\rangle_{2}$ & $|V\rangle_{2}$& $|H\rangle_{3}$ &$|V\rangle_{3}$\\\hline
$|H\rangle_{4}$ &  & $\bigcirc$ & & $\bigcirc$\\\hline
$|V\rangle_{4}$ & $\bigtriangleup$ &  & $\bigtriangleup$ & \\\hline
\end{tabular}
\caption{$\bigcirc$ and $\bigtriangleup$ represent the coupling to the first
and the second qubus beam, respectively. The XPM pattern in Merging gate is
that the first qubus beam is coupled to $|H\rangle_{4}$ of the ancilla photon
and $|V\rangle_{2,3}$ of the second photon after PBS, while the second beam to
$|V\rangle_{4}$ of the ancilla photon and $|H\rangle_{2,3}$ of the second
photon. }%
\label{tb4}%
\end{table}
The total state, $\left\vert \psi\right\rangle_{in} \left\vert
+\right\rangle_4 \left\vert \alpha\right\rangle \left\vert \alpha\right\rangle $, will be transformed to
\begin{align}
&  \frac{1}{\sqrt{2}}\left(  a\left\vert HHH\right\rangle _{124}+b\left\vert
HVV\right\rangle _{124}\right)  \left\vert \alpha\right\rangle \left\vert
\alpha\right\rangle \nonumber\\
&  +\frac{1}{\sqrt{2}}\left(  c\left\vert VHH\right\rangle _{134}+d\left\vert
VVV\right\rangle _{134}\right)  \left\vert \alpha\right\rangle \left\vert
\alpha\right\rangle \nonumber\\
&  +\frac{1}{\sqrt{2}}\left(  a\left\vert HHV\right\rangle _{124}+c\left\vert
VHV\right\rangle _{134}\right)  \left\vert \alpha e^{-i\theta}\right\rangle
\left\vert \alpha e^{i\theta}\right\rangle \nonumber\\
&  +\frac{1}{\sqrt{2}}\left(  b\left\vert HVH\right\rangle _{124}+d\left\vert
VVH\right\rangle _{134}\right)  \left\vert \alpha e^{i\theta}\right\rangle
\left\vert \alpha e^{-i\theta}\right\rangle
\end{align}
by the Entangler.
A bit flip $\sigma_{x}$ and a phase shifter
$\pi$ conditioned on the number-resolving detection results by a QND module will yield the following state
\begin{equation}
a\left\vert HHH\right\rangle _{124}+b\left\vert HVV\right\rangle
_{124}+c\left\vert VHH\right\rangle _{134}+d\left\vert VVV\right\rangle
_{134}.
\end{equation}
This entangled state of three photons can be rewritten as
\begin{align}
&  \frac{1}{\sqrt{2}}a|H\rangle_{1}(|+\rangle_{2}+|-\rangle_{2})|H\rangle
_{4}+\frac{1}{\sqrt{2}}b|H\rangle_{1}(|+\rangle_{2}-|-\rangle_{2}%
)|V\rangle_{4}\nonumber\\
&  +\frac{1}{\sqrt{2}}c|V\rangle_{1}(|+\rangle_{3}+|-\rangle_{3})|H\rangle
_{4}+\frac{1}{\sqrt{2}}d|V\rangle_{1}(|+\rangle_{3}-|-\rangle_{3}%
)|V\rangle_{4}.
\end{align}
After the interference of the single photon modes on path 2 and 3
\begin{align}
|\pm\rangle
_{2} & \rightarrow\frac{1}{\sqrt{2}}(|\pm\rangle_{2}+|\pm\rangle_{3}),\nonumber\\
|\pm\rangle_{3} &\rightarrow\frac{1}{\sqrt{2}}(|\pm\rangle_{2}-|\pm\rangle
_{3})
\end{align}
through a 50:50 BS, two PBS$_{\pm}$ send the single photon to four different paths numbered from $5$ to $8$,
giving the state
\begin{align}
&  \frac{1}{2}\left(  a\left\vert HH\right\rangle +b\left\vert HV\right\rangle
+c\left\vert VH\right\rangle +d\left\vert VV\right\rangle \right)
_{14}\left\vert +\right\rangle _{5}\nonumber\\
&  +\frac{1}{2}\left(  a\left\vert HH\right\rangle -b\left\vert
HV\right\rangle +c\left\vert VH\right\rangle -d\left\vert VV\right\rangle
\right)  _{14}\left\vert -\right\rangle _{6}\nonumber\\
&  +\frac{1}{2}\left(  a\left\vert HH\right\rangle +b\left\vert
HV\right\rangle -c\left\vert VH\right\rangle -d\left\vert VV\right\rangle
\right)  _{14}\left\vert +\right\rangle _{7}\nonumber\\
&  +\frac{1}{2}\left(  a\left\vert HH\right\rangle -b\left\vert
HV\right\rangle -c\left\vert VH\right\rangle +d\left\vert VV\right\rangle
\right)  _{14}\left\vert -\right\rangle _{8}.
\label{merge-out}
\end{align}
At this step, we can use a QND module on each path to project out the target state. In these QND modules the photon number non-resolving detectors can be simple APDs which output the same signal without telling the difference of the input states.
Finally, the classically feed-forwarded QND detection results control the operation $\sigma_{z}$ on path 1 and 4 to tailor the phases of the projected-out double photon state. The detected photon (in the state $|\pm\rangle$) on one of the paths can be used again in the next Merging gate.

\subsection{Two-qubit gate}
\label{2-qubit}

With a pair of C-path and Merging gate, it is very convenient to realize the two-qubit gate operations in the form $|H\rangle\langle H|\otimes U_1+|V\rangle\langle V|\otimes U_2$. If $U_1=I$ and $U_2=\sigma_x$, for instance, this gate will be a CNOT gate. Here we directly apply $U_1$ and $U_2$ on path 1 and 2 for the state $|\phi\rangle$ in Eq. (\ref{c-path}), and then the merging of the modes on path 1 and 2 will finish the transformation.

\begin{figure}[ptb]
\includegraphics[width=8.6cm]{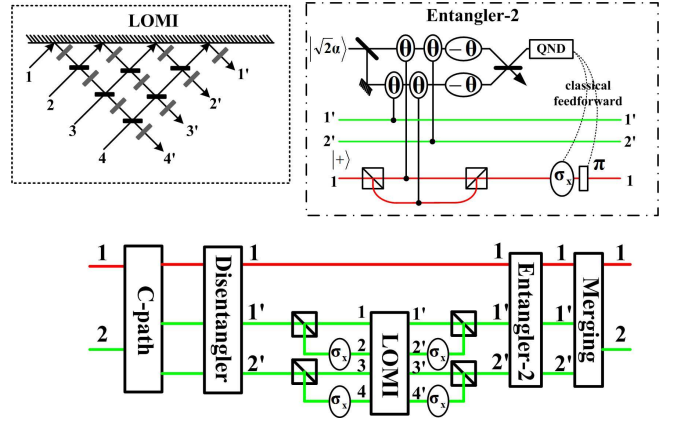}
\caption{(Color online) Schematic setup for general two-qubit gate. By a
C-path gate in Fig. \ref{c-path-p} and an Disentangler in Fig. \ref{two-photon},
the input two-photon states are transformed to the products single photon states, with the second photon on path $1'$ and $2'$ inheriting the structures of the input bi-photon states. Then the second photon becomes a qudit in four spatial modes under two
PBSs. Two $\sigma_x$ operations on path 2 and 4 make the same polarization for all spatial modes.
An LOMI in the dashed line follows up to implement the gate operation $U(4)$ on the single photon qudit. The inverse inverse transformation for the second photon back to polarization space relies on Entangler-2 in the
dash-dotted line, where the qubus beams interact with the photonic modes in the way slightly different from that in Entangler of Fig. \ref{merge}.
Finally, a Merging gate merges the spatial modes of the second photon, finishing a general two-qubit operation $U(4)$.}
\label{two-qubit-gate}
\end{figure}

The main point in this subsection is the realization for a general two-qubit gate $U\in U(4)$.
As has been proved early, three CNOT gates (together with the proper one-qubit operations) should be necessary in constructing a general two-qubit gate \cite{cnot, TD}. It looks that three pairs of C-path and Merging involving six number-resolving detections should be used to realize such two-qubit gate. However, fewer resources would be necessary if we work with the design in Fig. \ref{two-qubit-gate}.

In Fig. \ref{two-qubit-gate}, we first use a two-photon transformation to
convert an initial state $\left\vert \psi\right\rangle _{in}$ in Eq. (\ref{in}) to the form in Eq. (\ref{2-trans}).
Then, two PBS on the path $1^{\prime}$ and $2^{\prime}$ followed by two $\sigma_{x}$ operations on path 2 and 4 achieve the state
\begin{align}
 \left\vert +\right\rangle _{1}|\varphi\rangle&=\left\vert +\right\rangle _{1}\left(  a\left\vert H\right\rangle
_{1}+b\left\vert H\right\rangle _{2}+c\left\vert H\right\rangle _{3}%
+d\left\vert H\right\rangle _{4}\right)  \nonumber\\
&  \equiv\left\vert +\right\rangle _{1}\left(  a\left\vert 1\right\rangle
+b\left\vert 2\right\rangle +c\left\vert 3\right\rangle +d\left\vert
4\right\rangle \right)  ,
\end{align}
where $\left\vert 1\right\rangle ,\left\vert 2\right\rangle ,\left\vert
3\right\rangle ,\left\vert 4\right\rangle $ denote the spatial modes for the single photon.
Now the second photon is a qudit with 4 spatial modes.
Any operation $U\in U(4)$ can be implemented on this single photon qudit by a linear optical multi-port interferometer (LOMI)
in the dashed line of Fig. \ref{two-qubit-gate} \cite{Reck}. The LOMI transforms the qudit to
\begin{equation}
U|\varphi\rangle= a^{^{\prime}}\left\vert 1\right\rangle
+b^{^{\prime}}\left\vert 2\right\rangle +c^{^{\prime}}\left\vert
3\right\rangle +d^{^{\prime}}\left\vert 4\right\rangle  .
\end{equation}

Next, we should transform the single photon qudit back to polarization space.
By the inverse operations of $\sigma_{x}$ and two PBS, we have the following state
\begin{equation}
\left\vert +\right\rangle _{1}\left(  a^{^{\prime}}\left\vert H\right\rangle
_{1^{\prime}}+b^{^{\prime}}\left\vert V\right\rangle _{1^{\prime}}%
+c^{^{\prime}}\left\vert H\right\rangle _{2^{\prime}}+d^{^{\prime}}\left\vert
V\right\rangle _{2^{\prime}}\right)  .
\end{equation}
The conversion follows with a setup called Entangler-2 in the dash-dotted line of Fig. \ref{two-qubit-gate}.
Entangler-2 is a little bit different from an Entangler in Fig. \ref{merge}, with the first (second)
qubus beam coupled to $\left\vert H\right\rangle _{1}$ and $\left\vert
H/V\right\rangle _{2^{\prime}}$ ($\left\vert V\right\rangle _{2}$ and
$\left\vert H/V\right\rangle _{1^{\prime}}$).
Through such XPM processes in the Entangler-2, the following state can be obtained:
\begin{align}
&  \frac{1}{\sqrt{2}}\left(  a^{^{\prime}}\left\vert HH\right\rangle
_{1,1^{\prime}}+b^{^{\prime}}\left\vert HV\right\rangle _{1,1^{\prime}%
}\right)  \left\vert \alpha\right\rangle \left\vert \alpha\right\rangle
\nonumber\\
&  +\frac{1}{\sqrt{2}}\left(  c^{^{\prime}}\left\vert VH\right\rangle
_{1,2^{\prime}}+d^{^{\prime}}\left\vert VV\right\rangle _{1,2^{\prime}%
}\right)  \left\vert \alpha\right\rangle \left\vert \alpha\right\rangle
\nonumber\\
&  +\frac{1}{\sqrt{2}}\left(  a^{^{\prime}}\left\vert VH\right\rangle
_{1,1^{\prime}}+b^{^{\prime}}\left\vert VV\right\rangle _{1,1^{\prime}%
}\right)  \left\vert \alpha e^{-i\theta}\right\rangle \left\vert \alpha
e^{i\theta}\right\rangle \nonumber\\
&  +\frac{1}{\sqrt{2}}\left(  c^{^{\prime}}\left\vert HH\right\rangle
_{1,2^{\prime}}+d^{^{\prime}}\left\vert HV\right\rangle _{1,2^{\prime}%
}\right)  \left\vert \alpha e^{i\theta}\right\rangle \left\vert \alpha
e^{-i\theta}\right\rangle .
\end{align}
Then the photon number resolving detection by a QND module controls the indicated operations
on the photonic modes to achieve the state
\begin{equation}
a^{^{\prime}}\left\vert HH\right\rangle _{1,1^{\prime}}+b^{^{\prime}%
}\left\vert HV\right\rangle _{1,1^{\prime}}+c^{^{\prime}}\left\vert
VH\right\rangle _{1,2^{\prime}}+d^{^{\prime}}\left\vert VV\right\rangle
_{1,2^{\prime}}.
\end{equation}
It is then easy to obtain the final state,
\begin{equation}
|\psi\rangle_{out}=a^{^{\prime}}\left\vert HH\right\rangle +b^{^{\prime}}\left\vert
HV\right\rangle +c^{^{\prime}}\left\vert VH\right\rangle +d^{^{\prime}%
}\left\vert VV\right\rangle ,
\end{equation}
by merging the modes on path $1^{\prime}$ and $2^{\prime}$ with a Merging gate, realizing a general two-qubit operation $U\in U(4)$.

In this proposal, we first transform the two-photon state to a single photon qudit inheriting the coefficients of it,
and then perform the general operation $U\in U(4)$ on this single photon qudit by a linear optical circuit. Finally the tensor product of the qudit and the other single photon qubit will be transformed back to the target two-photon state by a modified Entangler and a Merging gate.
Compared with all other approaches thus far, this design uses less resources to realize a general two-qubit gate in deterministic way.

\subsection{General multi-qubit gate}
\label{m2}

In what follows, we will generalize the design of two-qubit gate to multi-qubit gate $U\in U(2^{n})$.
Firstly, we use a multi-photon transformation to obtain a corresponding
single photon qudit for the n-th photon as in Eq. (\ref{m-out}), and then perform the operation $U$ on
this single photon qudit by an LOMI with $2^{n}$ input and output ports.
Because the last single photon and the other $n-1$ photons will be in a tensor product state from Eq. (\ref{m-out}), the remaining work is to
entangle them to the proper output state of $U$. However, the last single photon has $2^{n-1}$ different spatial modes, so we should modify the Entangler in Fig. \ref{merge} to suit the case.

\begin{figure}[ptb]
\includegraphics[width=6.5cm]{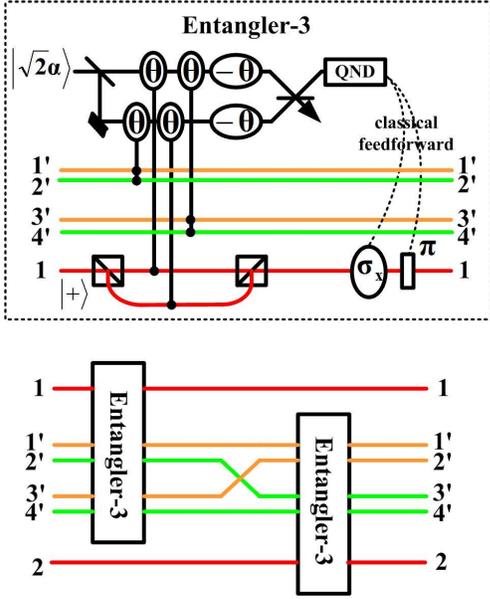}
\caption{(Color online) Schematic setup to entangle three photons as shown from Eq. (\ref{e1}) to Eq. (\ref{e3}).
The spatial modes of the third photon on path $1'$, $2'$, $3'$ and $4'$ are coupled to photon $1$ and photon $2$ in two Entangler-3 illustrated in the dashed line. The setup outputs an entangled triple-photon state in Eq. (\ref{e3}).}
\label{entangler-3}
\end{figure}

In triple-qubit gates, for example, the four spatial modes after the
transformation for the third photon to different spatial modes will be sent to
a modified Entangler called Entangler-3 (Fig. \ref{entangler-3}), where they
are divided into two parts ($1^{\prime},2^{\prime}$) and ($3^{\prime
},4^{\prime}$), having each part treated as one spatial mode like in an
Entangler-2. In this gate, the first (second) qubus beam is coupled to
both $\left\vert H\right\rangle _{1}$ and $\left\vert H/V\right\rangle _{3^{\prime
},4^{\prime}}$ ($\left\vert V\right\rangle _{2}$ and $\left\vert
H/V\right\rangle _{1^{\prime},2^{\prime}}$). After that, the
triple-photon state in Eq. (\ref{triple}) will be first transformed to
\begin{align}
& \frac{1}{\sqrt{2}}\left\vert +\right\rangle _{2}\left\vert H\right\rangle
_{1}\left(  \left\vert \phi_{1}\right\rangle _{1^{\prime}}+\left\vert \phi
_{2}\right\rangle _{2^{\prime}}\right)  \left\vert \alpha\right\rangle
\left\vert \alpha\right\rangle \nonumber\\
& +\frac{1}{\sqrt{2}}\left\vert +\right\rangle _{2}\left\vert V\right\rangle
_{1}\left(  \left\vert \phi_{3}\right\rangle _{3^{\prime}}+\left\vert \phi
_{4}\right\rangle _{4^{\prime}}\right)  \left\vert \alpha\right\rangle
\left\vert \alpha\right\rangle \nonumber\\
& +\frac{1}{\sqrt{2}}\left\vert +\right\rangle _{2}\left\vert V\right\rangle
_{1}\left(  \left\vert \phi_{1}\right\rangle _{1^{\prime}}+\left\vert \phi
_{2}\right\rangle _{2^{\prime}}\right)  \left\vert \alpha e^{-i\theta
}\right\rangle \left\vert \alpha e^{i\theta}\right\rangle \nonumber\\
& +\frac{1}{\sqrt{2}}\left\vert +\right\rangle _{2}\left\vert H\right\rangle
_{1}\left(  \left\vert \phi_{3}\right\rangle _{3^{\prime}}+\left\vert \phi
_{4}\right\rangle _{4^{\prime}}\right)  \left\vert \alpha e^{i\theta
}\right\rangle \left\vert \alpha e^{-i\theta}\right\rangle ,
\label{e1}
\end{align}
together with two qubus beams, and then to
\begin{equation}
\left\vert +\right\rangle _{2}\left(  \left\vert H\right\rangle _{1}\left\vert
\phi_{1}\right\rangle _{1^{\prime}}+\left\vert H\right\rangle _{1}\left\vert
\phi_{2}\right\rangle _{2^{\prime}}+\left\vert V\right\rangle _{1}\left\vert
\phi_{3}\right\rangle _{3^{\prime}}+\left\vert V\right\rangle _{1}\left\vert
\phi_{4}\right\rangle _{4^{\prime}}\right)
\label{e2}
\end{equation}
by post-selection with photon number resolving detection as in parity, C-path and Merging gate.
Next, the spatial modes of the third photon will be regrouped as in the lower
part of Fig. \ref{entangler-3}---it is to divide the four spatial modes into
two parts ($1^{\prime},3^{\prime}$) and ($2^{\prime},4^{\prime}$), and send
the above state to the second Entangler-3. The following process in the second Entangler-3,
\begin{align}
& \frac{1}{\sqrt{2}}\left\vert H\right\rangle _{2}\left(  \left\vert
H\right\rangle _{1}\left\vert \phi_{1}\right\rangle _{1^{\prime}}+\left\vert
V\right\rangle _{1}\left\vert \phi_{3}\right\rangle _{3^{\prime}}\right)
\left\vert \alpha\right\rangle \left\vert \alpha\right\rangle \nonumber\\
& +\frac{1}{\sqrt{2}}\left\vert V\right\rangle _{2}\left(  \left\vert
H\right\rangle _{1}\left\vert \phi_{2}\right\rangle _{2^{\prime}}+\left\vert
V\right\rangle _{1}\left\vert \phi_{4}\right\rangle _{4^{\prime}}\right)
\left\vert \alpha\right\rangle \left\vert \alpha\right\rangle \nonumber\\
& +\frac{1}{\sqrt{2}}\left\vert V\right\rangle _{2}\left(  \left\vert
H\right\rangle _{1}\left\vert \phi_{1}\right\rangle _{1^{\prime}}+\left\vert
V\right\rangle _{1}\left\vert \phi_{3}\right\rangle _{3^{\prime}}\right)
\left\vert \alpha e^{-i\theta}\right\rangle \left\vert \alpha e^{i\theta
}\right\rangle \nonumber\\
& +\frac{1}{\sqrt{2}}\left\vert H\right\rangle _{2}\left(  \left\vert
H\right\rangle _{1}\left\vert \phi_{2}\right\rangle _{2^{\prime}}+\left\vert
V\right\rangle _{1}\left\vert \phi_{4}\right\rangle _{4^{\prime}}\right)
\left\vert \alpha e^{i\theta}\right\rangle \left\vert \alpha e^{-i\theta
}\right\rangle \nonumber\\
& \rightarrow\left(  \left\vert HH\right\rangle \left\vert \phi_{1}%
\right\rangle _{1^{\prime}}+\left\vert HV\right\rangle \left\vert \phi
_{2}\right\rangle _{2^{\prime}}\right)  _{123}\nonumber\\
& +\left(  \left\vert VH\right\rangle \left\vert \phi_{3}\right\rangle
_{3^{\prime}}+\left\vert VV\right\rangle \left\vert \phi_{4}\right\rangle
_{4^{\prime}}\right)  _{123},
\label{e3}
\end{align}
can be realized too. For a general multi-photon state, we will just apply this procedure to
the spatial modes of the $n$-th photon iteratively. The multi-photon state in Eq. (\ref{m-out}) will be transformed correspondingly to
\begin{equation}
\left(  \left\vert H\cdots H\right\rangle \left\vert \phi_{1}\right\rangle
_{1^{\prime}}+\cdots+\left\vert V\cdots V\right\rangle \left\vert
\phi_{2^{n-1}}\right\rangle _{2^{n-1\prime}}\right)  _{1\cdots n},
\end{equation}
an entangled state of $n$ photons.

\begin{figure}[ptb]
\includegraphics[width=7.7cm]{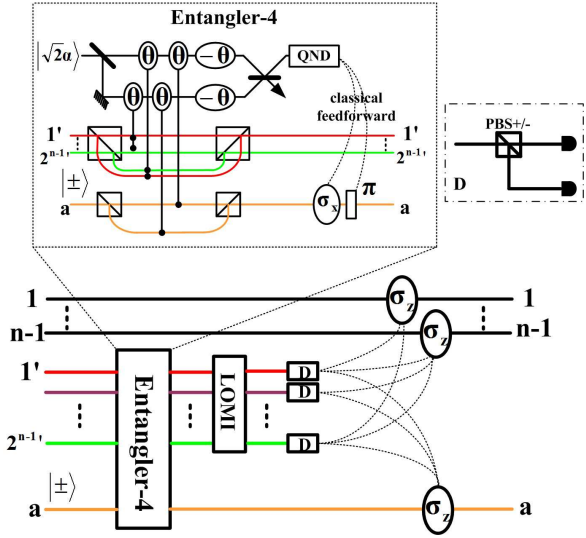}
\caption{(Color online) General Merging-n gate.
In Entangler-4 within the dashed line, the two qubus beams should be coupled to all spatial modes of the n-th photon as indicated.
Entangler-4 is used to entangle the n-th photon in multiple spatial modes with the ancilla
photon. An LOMI will then perform the QFT in \ref{qft} or other possible multi-rail transformations, which are followed by the detections D in the dash-dotted line. Their detection results will be classically
feed-forwarded to redress the phases of the involved photonic modes. The input states with the n-th photon in multiple spatial modes will then be merged to the spatial modes of the other $n-1$ photons and the ancilla photon a.}
\label{merge-n}
\end{figure}

Then, similar to a two-qubit gate, we should merge the n-th photon in multiple spatial modes to only one spatial mode without changing anything else. However, the standard Merging gate is not suitable to do the work as there are more than two spatial modes, so we should generalize it to the one
called Merging-n gate shown in Fig. \ref{merge-n}. In this gate, an ancilla single photon $\left\vert +\right\rangle
_{a}$ is used, and the last photon and the ancilla in the state $\left\vert +\right\rangle _{a}$\ are processed by a setup called
Entangler-4, where the first (second) qubus beam is coupled to
$\left\vert
H\right\rangle _{a}$ and $\left\vert V\right\rangle _{1^{\prime}%
\cdots2^{n-1\prime}}$ ($\left\vert V\right\rangle _{a}$ and $\left\vert
H\right\rangle _{1^{\prime}\cdots2^{n-1\prime}}$) simultaneously.
The Entangler-4 outputs the following state ($a$ denotes the ancilla):
\begin{equation}
\left(  \left\vert H\cdots H\right\rangle \left\vert \phi_{1}^{^{\prime}%
}\right\rangle _{1,a}+\cdots+\left\vert V\cdots V\right\rangle \left\vert
\phi_{2^{n-1}}^{^{\prime}}\right\rangle _{2^{n-1},a}\right)  _{1\cdots n},
\label{out-m}
\end{equation}
where $\left\vert \phi_{j}^{^{\prime}}\right\rangle _{j,a}=\left(  \alpha
_{j}\left\vert HH\right\rangle +\beta_{j}\left\vert VV\right\rangle \right)
_{j,a}$, for $j=1,\cdots,2^{n-1}$.
The difference in the next step from a standard Merging gate is that a circuit performing quantum Fourier transform (QFT) \cite{Nielsen} ($\left\vert j\right\rangle ,\left\vert k\right\rangle $ denote the
spatial modes, and $N=2^{n-1}$),
\begin{equation}
\left\vert j\right\rangle =\frac{1}{\sqrt{N}}\overset{N-1}{\underset
{k=0}{{\sum}}}e^{2\pi ijk/N}\left\vert k\right\rangle ,
\label{qft}
\end{equation}
instead of only one 50:50 BS in a standard Merging gate should be used for the interference of the spatial modes of the n-th photon.
The QFT for each of $2^{n-1}$ modes of the n-th photon
in Eq. (\ref{out-m}) is
\begin{align}
\left\vert \phi_{j}^{^{\prime}}\right\rangle _{j,a} &  =\frac{1}{\sqrt{2}%
}\left(  \left\vert \phi_{j}\right\rangle _{a}\left\vert +\right\rangle
_{j}+\sigma_{z}\left\vert \phi_{j}\right\rangle _{a}\left\vert -\right\rangle
_{j}\right)  \nonumber\\
&  \rightarrow\frac{1}{\sqrt{2N}}\left(  \left\vert \phi_{j}\right\rangle
_{a}\overset{N-1}{\underset{k=0}{{\sum}}}e^{2\pi ijk/N}\left\vert
+\right\rangle _{k}\right.  \nonumber\\
&  \left.  +\sigma_{z}\left\vert \phi_{j}\right\rangle _{a}\overset
{N-1}{\underset{k=0}{{\sum}}}e^{2\pi ijk/N}\left\vert -\right\rangle
_{k}\right)  .
\end{align}
By a PBS$_{\pm}$ on each output spatial mode and the corresponding conditional
operations involving $\sigma_{z}$ and phase shifts $e^{2\pi
ijk/N}$, the desired output state will be projected out from Eq. (\ref{out-m}).
Of course, simpler circuits could be found to replace the QFT circuits.
For example, in the case of three-photon gates,
\begin{equation}
U=\frac{1}{2}\left(
\begin{array}
[c]{cccc}%
1 & 1 & 1 & 1\\
1 & 1 & -1 & -1\\
1 & -1 & -1 & 1\\
1 & -1 & 1 & -1
\end{array}
\right) ,
\end{equation}
would perform the interference of the spatial modes of the last photon as well as a QFT circuit.
Then, the remaining conditional operations are just $\sigma_{z}$ without any phase shift.

Merging-n gate will reduce to a standard Merging gate if $n=2$. With the gate, an arbitrary multi-qubit gate $U\in U(2^{n})$ could be realized deterministically. Here $n-1$ C-path gates (one C-path and $n-2$ modified C-path gates), $n$ Entanglers ($n-1$ Entangler-3 and one
Entangler-4), one ancilla single photon and plus two LOMI
(a $2^{n}$-dimensional one for the gate unitary operation and a $2^{n-1}$-dimensional one for QFT) are the resources to realize the gate.
A significant advantage of this proposal is that it works for any circuit in principle even if we do not know its decomposition into two-qubit gates.

\subsection{Multi-control gate}
Although it is possible to realize an n-qubit gate $U\in U(2^{n})$ with the current technology as we apply the above-discussed transformations approaches, the necessary resources (BS and phase shifters) for the LOMIs in the setups will exponentially grow with $n$. The number of the modes to be coupled to the qubus beams in a C-path-2 and an Entangler-4 is also exponentially large with $n$. It is therefore not realistic to apply the method to build a gate of large $n$. However, the resources for constructing a special class of gates---multi-control gates---can be greatly reduced if we apply C-path and Merging gates. The operation of this type of gates is given as
\begin{align}
C^n(U_1)&=(I\otimes\cdots I-|V\cdots V\rangle\langle V\cdots V|)\otimes I\nonumber\\
&+|V\cdots V\rangle\langle V\cdots V|\otimes U_1.
\label{control}
\end{align}
It effects the action of $U_1$ on the last photon under the control of $|V\rangle$ component of the remaining $n-1$ photons.
The gate is a key multi-qubit gate called Toffoli gate \cite{Toffoli, B} for $U_1=\sigma_x$.

\begin{table}
\begin{tabular}
[c]{|c|c|c|}\hline
& $|H/V\rangle_{4}$ & $|H/V\rangle_{5}$\\\hline
$|H\rangle_{2}$ &  & $\bigtriangleup$\\\hline
$|V\rangle_{2'}$ &  & $\bigtriangleup$\\\hline
$|H\rangle_{3}$ &  &  $\bigtriangleup$\\\hline
$|V\rangle_{3}$ & $\bigcirc$ & \\\hline
\end{tabular}
\caption{$\bigcirc$ and $\bigtriangleup$ represent the coupling to the first
and the second qubus beam, respectively. The XPM pattern in such C-path-3 gate
is that the first qubus beam is coupled to $|V\rangle_{3}$ of the control
photon $C_{2}$ and all modes of the target photon $T$\ on path 4 after BS,
while the second beam to all $|H\rangle_{3}$,  $|H\rangle_{2}$ and  $|V\rangle_{2'}$ (which becomes $|H\rangle_{2'}$ after the bit flip) of the second control photon
$C_{2}$ in Fig. \ref{Toffoli}, and
those on path 5 from the target photon $T$. }%
\label{tb5}%
\end{table}

\begin{figure}[ptb]
\includegraphics[width=6cm]{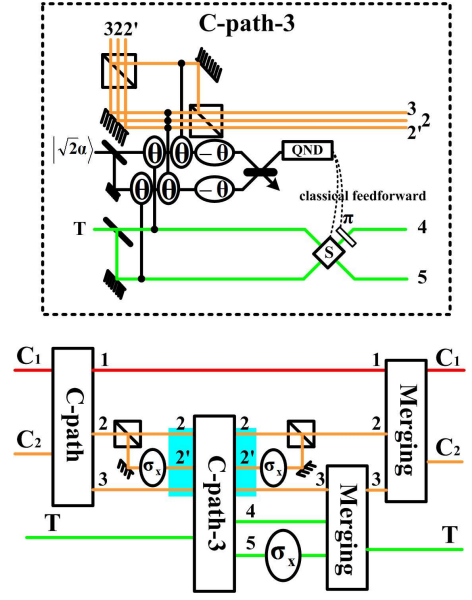}
\caption{(Color online) Triple-photon Toffoli gate. Here one C-path gate and one C-path-3 gate, which is illustrated in dashed line, transform every photon to two spatial modes. The qubus interaction pattern in C-path-3 is summarized in Tab. \ref{tb5} for convenience.
The target photon will be on path 5 conditioned on $|V\rangle$ of both two control photons. Here two $\sigma_x$ are performed on path $2'$ in order to arrange the photonic modes as in C-path-3 gate. A $\sigma_x$ operation on path 5 and two following Merging gates will then complete the Toffoli gate operation. All bit flip $\sigma_x$ here can be performed by $\lambda/2$ wave plate.}
\label{Toffoli}
\end{figure}

A Toffoli gate is a universal element in quantum computation, i. e., any quantum network can be constructed with this multi-qubit gate and the proper single-qubit gates. The gate is useful in implementing Shor's algorithm
\cite{Shor} and quantum error correction \cite{G}. Several recent works have been attempted to optical realization of this gate \cite{F, R, L, compact, Lin}. In most of previous approaches, Toffoli gate is constructed by means of decomposing the gate into two-qubit gates. There should be at least five two-qubit gates in constructing a triple-qubit Toffoli gate \cite{SD}, and a general Toffoli gate of $n$ qubits requires $\cal{O}$ $(n^2)$ two-qubit gates \cite{B}. It was proved that the resources could be reduced with part of input state encoded as a qudit in operation \cite{R}. Here we show in what follows that a deterministic optical Toffoli gate, which uses the resources increasing linearly with the size of an input, is possible even if the input is simply encoded as a multi-qubit product state in Eqs. (\ref{1}) and (\ref{20}).

\begin{figure}[ptb]
\includegraphics[width=5.5cm]{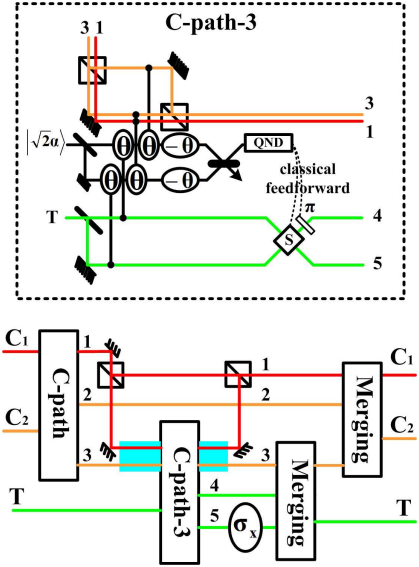}
\caption{(Color online) Alternative design for triple-photon Toffoli gate.
The difference from the design in Fig. \ref{Toffoli} is the different coupling pattern in C-path-3 gate. This design saves one mode to be coupled to a qubus beam.}
\label{Toffoli-2}
\end{figure}

In Fig. \ref{Toffoli} and \ref{Toffoli-2}, we outline the designs to realize such gate with weak nonlinearity and linear optics. Like a general multi-qubit gate in \ref{m2}, these designs do not rely on the decomposition into two-qubit gates. Their difference from a general multi-qubit gate
is no transformation of one input photon to $2^n$ spatial modes. The idea is that the first control photon sends the second photon to two different paths conditioned on its polarizations, and the modes on the second path of the second photon continue to control that of the third photon, and so forth to the last photon. This is what happens to an $n$-control Toffoli gate realizing Eq. (\ref{control}). The number of the spatial modes for each photon, including the target photon, will be two after such operations, so $U_1$ could be performed on one spatial mode of the target to realize the gate.

We illustrate the realization of Toffoli gate with the design in Fig. \ref{Toffoli}. The gate processes any input like the following with each qubit
encoded in the respective polarization space:
\begin{align}
|\Psi\rangle_{C_1C_2T}  &  =A_{1}\left\vert HHH\right\rangle
+A_{2}\left\vert HHV\right\rangle +A_{3}\left\vert HVH\right\rangle
\nonumber\\
&  +A_{4}\left\vert HVV\right\rangle +A_{5}\left\vert VHH\right\rangle
+A_{6}\left\vert VHV\right\rangle \nonumber\\
&  +A_{7}\left\vert VVH\right\rangle +A_{8}\left\vert VVV\right\rangle.
\end{align}
Firstly, the polarizations of the photon $C_1$ control the second photon to two different paths $2$ and $3$ through a C-path gate, giving the state
\begin{align}
|\Phi\rangle &=(A_{1}\left\vert HHH\right\rangle +A_{2}\left\vert HHV\right\rangle+A_{3}\left\vert HVH\right\rangle \nonumber\\
&+ A_{4}\left\vert HVV\right\rangle)_{12T}+(A_{5}\left\vert VHH\right\rangle +A_{6}\left\vert VHV\right\rangle \nonumber\\
&+A_{7}\left\vert VVH\right\rangle +A_{8}\left\vert VVV\right\rangle )
_{13T} \label{Tf-1}.
\end{align}
At this time, if the modes of the second photon on path 3 are to control the paths of the target photon through a standard C-path gate, there will be a problem of how to arrange the coupling of the qubus beams with all terms in Eq. (\ref{Tf-1}). The qubus beams are in tensor product state with all single photon modes, but the modes $|H/V\rangle_3$ on path 3 are only part of the superposition. We should consider the coupling of two qubus beams with the other modes in the superposition of Eq. (\ref{Tf-1}). Therefore, we should modify the standard C-path gate to that in dashed line of Fig. \ref{Toffoli}. We call it C-path-3 gate to distinguish it from the prevously discussed C-path and C-path-2 gate.
A 50:50 BS divides the target photon into two paths $4$ and $5$ in Fig. \ref{Toffoli}, and then two qubus beams interact with the corresponding photonic modes as summarized in Tab. \ref{tb5}, realizing the following transformation
\begin{widetext}
\begin{align}
|\Phi\rangle |\alpha\rangle|\alpha\rangle &\rightarrow \frac{1}{\sqrt{2}}|HH\rangle_{12}\{ (A_1|H\rangle_4+|A_2|V\rangle_4)|\alpha\rangle|\alpha\rangle+(A_1|H\rangle_5+|A_2|V\rangle_5)|\alpha e^{-i\theta}\rangle|\alpha e^{i\theta}\rangle\}
\nonumber\\
&+\frac{1}{\sqrt{2}}|HV\rangle_{12'}\{ (A_3|H\rangle_4+|A_4|V\rangle_4)|\alpha\rangle|\alpha\rangle+(A_3|H\rangle_5+|A_4|V\rangle_5)|\alpha e^{-i\theta}\rangle|\alpha e^{i\theta}\rangle\}\nonumber\\
&+\frac{1}{\sqrt{2}}|VH\rangle_{13}\{ (A_5|H\rangle_4+|A_6|V\rangle_4)|\alpha\rangle|\alpha\rangle+(A_5|H\rangle_5+|A_6|V\rangle_5)|\alpha e^{-i\theta}\rangle|\alpha e^{i\theta}\rangle\}\nonumber\\
&+\frac{1}{\sqrt{2}}|VV\rangle_{13}\{ (A_7|H\rangle_4+|A_8|V\rangle_4)|\alpha e^{i\theta}\rangle|\alpha e^{-i\theta}\rangle+(A_7|H\rangle_5+|A_8|V\rangle_5)|\alpha \rangle|\alpha \rangle\},
\end{align}
\end{widetext}
together with the phase shift $-\theta$ on two qubus beams. Here the modes of the second photon on path 2 are divided into $|H\rangle_2$ and $|V\rangle_{2'}$ by a PBS. A triple-phootn state
\begin{align}
&  \left\vert H\right\rangle _{1}\left(  A_{1}\left\vert HH\right\rangle
+A_{2}\left\vert HV\right\rangle +A_{3}\left\vert VH\right\rangle
+A_{4}\left\vert VV\right\rangle \right)  _{24}\nonumber\\
&  +\left\vert VH\right\rangle _{13}\left(  A_{5}\left\vert H\right\rangle
+A_{6}\left\vert V\right\rangle \right)  _{4}+\left\vert VV\right\rangle
_{13}\left(  A_{7}\left\vert H\right\rangle +A_{8}\left\vert V\right\rangle
\right)  _{5}
\end{align}
will be post-selected out by a similar procedure to that in a standard C-path gate.
Here the target photon passes through path 5 only if both control
photons are in the state $\left\vert V\right\rangle $. A bit flip
$\sigma_{x}$ on path $5$ alone will then finish the target operation.
The remaning steps are the merging of the modes on path 4, 5 and on path 2, 3, respectively.
They are easy to perform with two standard Merging gates. A triple-photon Toffoli gate implementing the map,
\begin{align}
|\Psi\rangle_{C_1C_2T} &\rightarrow A_{1}\left\vert HHH\right\rangle
+A_{2}\left\vert HHV\right\rangle +A_{3}\left\vert HVH\right\rangle
\nonumber\\
&  +A_{4}\left\vert HVV\right\rangle +A_{5}\left\vert VHH\right\rangle
+A_{6}\left\vert VHV\right\rangle \nonumber\\
&  +A_{7}\left\vert VVV\right\rangle +A_{8}\left\vert VVH\right\rangle,
\end{align}
will be thus realized.

The design for the Toffoli gate in this approach is not unique. The other layout in Fig. \ref{Toffoli-2} realizes an optical Toffoli gate as well, but the interaction pattern for the qubus beams in C-path-3 gate is different from that in Fig. \ref{Toffoli}--- the second beam interacts with the modes of both first and second photon to save one mode for coupling.

\begin{figure}[ptb]
\includegraphics[width=8.0cm]{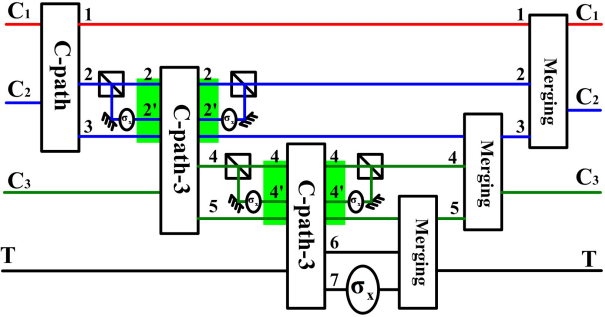}
\caption{(Color online) Schematic setup for triple-control or four-photon Toffoli
gate. It is a simple generalization from Fig. \ref{Toffoli}. Since the ancilla photon for the Merging gates can be recycled, a general Toffoli gate only needs one ancilla photon in principle.}
\label{toffoli-4}
\end{figure}

To construct a general Toffoli gate with $n$ control photons, we should apply $n-1$ C-path-3 gates, togther with one standard C-path gate.
The interaction pattern in Tab. \ref{tb5} for the $(k-1)$-th C-path-3 gate (where the modes of the $k$-th photon on the second spatial path control the paths of the $k+1$-th photon) will be simply generalized with the second beam coupled the second spatial mode of the $k+1$-th photon as well as with the $|H\rangle$ mode of the $k$-th photon on the second path and both $|H\rangle$ and $|V\rangle$ mode of the $k$-th photon on the first path. Meanwhile, the first qubus beam should be coupled to the first spatial mode of the $k+1$-th photon as well as the $|V\rangle$ mode
of the $k$-th photon on the second spatial path. In Fig. \ref{toffoli-4}, we give the schematic setup of triple-control Toffoli gate, for example.
The circuit resources for an $n$-control Toffoli gate are only $n$ pairs of C-path (C-path-3) and Merging gates. The design of an $n$-qubit Toffoli gate can be regarded as a simple generalization of that for CNOT gate (single-qubit control) mentioned at the beginning of
\ref{2-qubit}.

In addition, more general multi-control gate called $C^{n}(U_{k})$
gate \cite{Nielsen}, which performs the multi-qubit $U_{k}$ operation on $k$ target
photons controlled by $n$\ control photons, i.e.,
\begin{align}
C^{n}(U_{k})  &  =(\underbrace{I\otimes\cdots I}\limits_{n}-|V\cdots V\rangle\langle V\cdots
V|)\otimes (\underbrace{I\otimes \cdots I}\limits_{k}) \nonumber\\
&  +|V\cdots V\rangle\langle V\cdots V|\otimes U_{k},
\end{align}
can be realized in a similar way.
We could combine the multi-control gate idea discussed here with the multi-photon
transformation and its inverse in \ref{m1} and \ref{m2}. The circuit resources are $n+k$ C-path gates
(one C-path gate for the first two control photons, $n-1$
C-path-3 gates for the next $n-1$\ control photons, $k$ C-path-3 gates for the
final control photon and $k$ target photons), the gates to transform k-photon state to a corresponding single photon and their inverse, the LOMI for $U_{k}$ on single photon, and $n+k$ Merging gates. All these element gates together could realize a $C^{n}(U_{k})$ gate deterministically.

\section{Conclusion}
We present the methods of transforming a class of multi-photon states to the corresponding single photon states which inherit the structures of the initial multi-photon states. These transformations make the deterministic unitary operations and POVMs on the multi-photon states in the form
of tensor products of multiple single photon qubits possible. Since the input states of various quantum information processing tasks can be encoded in such form, these transformations and their inverses, which are implemented by a finite number of element gates, would be found applications in the related fields. In this paper we have discussed the realization of multi-photon state discrimination and the construction of quantum logic gates in the transformation approach. We especially discuss the realization of parity gate and Toffoli gate, which have wide applications in quantum information technology. The technical ingredients for this circuit or network based approach are linear optics and weak non-linearity, 
which are within the current technology. The circuits used for the transformations may add to the tool kits
to process photonic states.

\begin{acknowledgments}
B. H. thanks C. F. Wildfeuer for helpful dicussions. This work is partially supported by a PSC-CUNY grant.
\end{acknowledgments}

\end{document}